\documentclass[aps,pra, reprint,groupedaddress]{revtex4-1}

\usepackage{graphicx}
\usepackage{amssymb}
\usepackage{epstopdf}

\usepackage{hyperref} 

\begin{document}

\title{Photoassociation to the $2\,^1\Sigma_g^+$ state in ultracold $^{85}$Rb$_2$ in the presence of a shape resonance}
\author{M. A. Bellos, R. Carollo, D. Rahmlow, J. Banerjee, E. E. Eyler, P. L. Gould, and W. C. Stwalley}
\affiliation{Department of Physics, University of Connecticut, Storrs, Connecticut 06269, USA}

\date{\today}

\begin{abstract}
We report the first observation of photoassociation to the $2\,^1\Sigma_g^+$ state of $^{85}$Rb$_2$. We have observed two vibrational levels ($v'$=98, 99) below the $5s_{1/2}+5p_{1/2}$ atomic limit and eleven vibrational levels ($v'$=102-112) above it. The photoassociation---and subsequent spontaneous emission---occur predominantly between 15 and 20 Bohr in a region of internuclear distance best described as a transition between Hund's case (a) and Hund's case (c) coupling. The presence of a $g$-wave shape resonance in the collision of two ground-state atoms affects the photoassociation rate and lineshape of the $J'$= 3 and 5 rotational levels.
\end{abstract}


\maketitle

\section{Introduction}

Photoassociation (PA) of ultracold atoms has become a standard tool for high-resolution spectroscopy \cite{stwalley99}, with a wide range of applications \cite{jones06}. There is strong interest in producing cold and ultracold molecules in their lowest electronic and rovibrational state. Molecules in their lowest energy state are immune to inelastic collisions and can be used as a platform for quantum information \cite{carr09} and cold chemistry \cite{krems08}. A wide variety of experiments target the formation of stable molecules at low temperatures, such as buffer gas cooling \cite{hutzler12}, Stark deceleration \cite{bethlem99},  STIRAP transfer of magnetoassociated atoms \cite{winkler07} and photoassociated atoms \cite{aikawa10}, direct laser cooling of molecules \cite{barry12}, and rotating supersonic sources \cite{gupta99}, to name a few.

Traditionally, photoassociation has been limited to vibrational levels near the asymptotic limit of the excited electronic state, and consequently to large internuclear separations. Photoassociation at short range was believed to have very low rates due to the small amplitude of the ground-state wavefunction at short internuclear distances. However, photoassociation of LiCs \cite{deiglmayr08} and more recently of Rb$_2$ \cite{bellos11}, RbCs \cite{gabbanini11,*rbcs2012} and NaCs \cite{zabawa11} has shown that short-range photoassociation is possible, allowing the formation of deeply-bound molecules, including the lowest rovibrational levels.

Several factors can increase the photoassociation rate at short range. The amplitude of the ground-state wavefunction can be increased at short range by scattering resonances, i.e., shape resonances or Feshbach resonances \cite{chin10}. The amplitude of the excited-state wavefunction can be large if the excited state is constrained over a narrow range of internuclear distances, e.g. at the bottom of a potential well, or, in quasibound levels where the outer turning points are at small distances. Furthermore, transition dipole moments can increase at short range, thereby increasing the transition rate, although this is not the case in this work.

In previous work, we photoassociated to the $1\,^3\Pi_g$ state of Rb$_2$ and observed spontaneous decay to the lowest vibrational levels of the $a\,^3\Sigma_u^+$ state \cite{bellos11}. In some favorable cases, the spontaneous decay populated mostly a single vibrational level, the $v''$=0 level.

In this work we photoassociate to the neighboring $2\,^1\Sigma_g^+$ state and again observe that transitions to high rotational levels yield stronger signals than to low rotational levels. We now attribute this atypical rotational distribution to the presence of a shape resonance. The $2\,^1\Sigma_g^+$ state is appealing because it contains both red-detuned (i.e. bound) levels and blue-detuned (i.e. quasibound) levels. Also, since these levels spontaneously decay to intermediate vibrational levels of the $a\,^3\Sigma_u^+$ state, one can now populate vibrational levels at the bottom, middle, and top \cite{lozeille06} of the $a\,^3\Sigma_u^+$ potential well, with relatively narrow vibrational distributions.

\section{Experimental Setup}

Our experiment is performed in a magneto-optical trap (MOT) of $^{85}$Rb with a peak density of $\sim 1\times 10^{11}$ cm$^{-3}$, a total atom number of $\sim 8 \times 10^7$, and a temperature of $\sim 120$ $\mu$K. We load the MOT from a getter source, with a loading time of $\sim 2$ s and a non-alkali background pressure $<1 \times 10^{-10}$ torr. The MOT trapping laser is locked 14 MHz below the $\left | 5s_{1/2}, F = 3 \right> \rightarrow \left | 5p_{3/2}, F' = 4 \right >$ atomic transition, and a repumping laser is tuned to the $\left | 5s_{1/2}, F = 2 \right> \rightarrow \left | 5p_{3/2}, F' = 3 \right >$ transition to prevent buildup of atoms in the lower $F = 2$ hyperfine ground level. The energy splitting between the two hyperfine ground levels is one that routinely appears in our spectra as the presence of ``hyperfine ghost'' lines. These ``hyperfine ghost'' lines occur 0.10126 cm$^{-1}$ above strong PA transitions (as shown in Fig. \ref{v109}, for example) and may originate from short-range hyperfine-changing collisions \cite{boesten96,boesten99}.

We photoassociate with a single-mode cw Ti:Sapphire laser (Coherent 899-29) pumped by an argon ion laser (Coherent Innova 400). After fiber optic coupling we have over 1 W of usable power at the vacuum chamber. The PA laser is weakly focused to the size of the MOT, yielding a maximum PA intensity of about 100 W/cm$^{2}$. The tuning range used for this experiment varied between 780 nm and 795 nm.  Resonantly enhanced multi-photon ionization (REMPI) is performed by use of a pulsed dye laser (Continuum ND6000). It is operated between 625 nm and 675 nm using a DCM dye solution. The pulsed dye laser is pumped at 10 Hz by a 532 nm Nd:YAG laser (Spectra-Physics Lab 150). The REMPI pulse energy and linewidth are 5 mJ and 0.5 cm$^{-1}$, respectively.

After ionization, molecules are detected by a discrete dynode multiplier (ETP model 14150). Two electric field grids focus the ions through a long field-free tube to the detector, resulting in a small detection region centered at the location of the MOT. Molecular ions are distinguished from scattered photons and atomic ions by time-of-flight mass spectrometry, and the signal is integrated by a gated boxcar integrator (SRS model SR250). We turn off the MOT lasers starting 10 $\mu$s before each ionizing pulse and ending 10 $\mu$s after it to depopulate the excited  $\left |5p_{3/2} \right>$ state so as to decrease Rb$^+$ signals.

\section{Molecule formation pathways and transition moments}

The $2\,^1\Sigma_g^+$ state has been previously observed by laser-induced fluorescence from the highly excited $C\,(2)\,^1\Pi_u$ state \cite{amiot87}. To the best of our knowledge the present work is the first observation of the $2\,^1\Sigma_g^+$ state through excitation rather than decay. This state has eluded many experiments because excitation from the ground $X\,^1\Sigma_g^+$ state is forbidden by single-photon electric dipole parity selection rules. Furthermore, single-photon excitation from deeply-bound levels of the $a\,^3\Sigma_u^+$ state is forbidden by spin selection rules. However, by photoassociating from the triplet state of colliding atoms at a range of internuclear distances where spin is not a good quantum number, the spin selection rule breaks down, allowing the transition.

\begin{figure}[]
\includegraphics[scale=0.75]{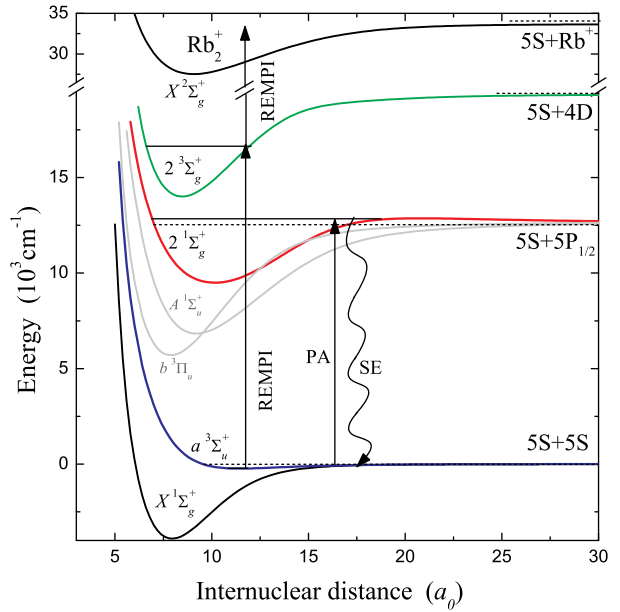}
\caption{\label{pec+pathway}(Color online) Potential energy curves with molecule formation and detection pathways. Arrows indicate photoassociation (PA) to $2\,^1\Sigma_g^+$, spontaneous emission (SE) to $a\,^3\Sigma_u^+$, and REMPI to Rb$_2^+$. Note that the $2\,^1\Sigma_g^+$ state has a barrier of 250 cm$^{-1}$ at $R=19$ $a_0$. Potential energy curves are from Refs. \cite{allouche12, jraij03}.}
\end{figure}

The photoassociation laser thus converts a small fraction of colliding atom pairs in the MOT into molecules in the $2\,^1\Sigma_g^+$ state. The subsequent decay of these excited molecules forms molecules in the $X\,^1\Sigma_g^+$ and $a\,^3\Sigma_u^+$ states through a variety of pathways, as discussed below. These $X$ or $a$ state molecules can then be detected by REMPI through any of several possible intermediate states. In this experiment we use REMPI through the $2\,^3\Sigma_g^+$ state to detect $a\,^3\Sigma_u^+$ molecules as shown in Fig. \ref{pec+pathway}. The REMPI laser monitors the population of a single vibrational level in the $a\,^3\Sigma_u^+$ state, typically between $v''$=18 and 24. We produce photoassociation spectra by scanning the PA laser while fixing the REMPI laser on resonance with an intermediate state.

\begin{figure}[]
\includegraphics[scale=0.8]{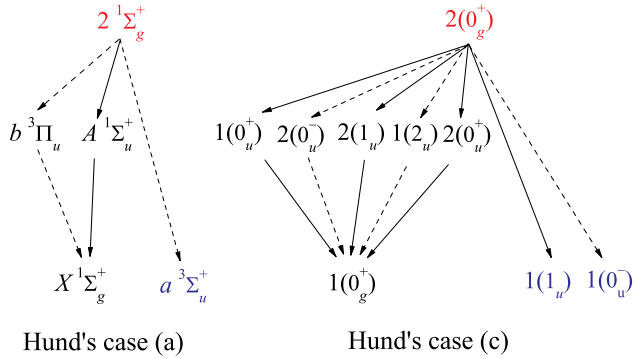}
\caption{\label{decay}(Color online) Some spontaneous emission pathways in Hund's case (a) and Hund's case (c) basis sets. Solid arrows denote allowed transitions and dashed arrows denote forbidden transitions according to one-photon E1 selection rules. Some transitions that are forbidden in Hund's case (a) are allowed in Hund's case (c).}
\end{figure}

It is useful to consider electric dipole (E1) selection rules for spontaneous emission from $2\,^1\Sigma_g^+$ to lower states using first the Hund's case (a) basis and then Hund's case (c) basis. The $2\,^1\Sigma_g^+$ state corresponds to the 2($0_g^+$) state in Hund's case (c) notation, which further correlates to the $5s_{1/2}$+$5p_{1/2}$ atomic limit \cite{stwalley12}. For Hund's case (a), single-photon decay from $2\,^1\Sigma_g^+$ to $X\,^1\Sigma_g^+$ is forbidden by parity selection rules ($g \leftrightarrow u$). However, a two-step cascade decay through the $A\,^1\Sigma_u^+$ state is allowed, and has been observed in Cs$_2$ \cite{viteau09}. The first step of such a cascade typically has a low transition rate due to the low transition frequency. This transition rate is given by summing over Einstein $A$ coefficients,
\begin{equation}
   A_{v'\rightarrow v''} \propto \nu^3 \, \bigl\vert\ \langle \psi_{v''} \mid \mu (R)\mid \psi_{v'} \rangle \, \bigr\vert^2,
\end{equation}
where $\psi_{v'}$ is the upper-state wavefunction, $\psi_{v''}$ is the lower-state wavefunction, $\mu (R)$ is the transition dipole moment as a function of internuclear distance, and $\nu$ is the transition frequency.

Decay to the $a\,^3\Sigma_u^+$ state is spin forbidden. As mentioned above, we nevertheless observe population in this state due to singlet-triplet mixing. This corresponds to Hund's case (c), where the spin quantum number is not well defined. Decay from 2($0_g^+$) to the ground 1($0_g^+$) state is still forbidden by the ($g \leftrightarrow u$) selection rule, just as in Hund's case (a). However decay from 2($0_g^+$) to 1($1_u$) (one of the two components of the $a\,^3\Sigma_u^+$ state, also denoted as $a\,^3\Sigma_u^+ (\Omega=1_u)$) is now allowed. Furthermore, the decay to 1($1_u$) dominates over other states, as it has the largest $\nu^3$ factor in Eq. 1.  These decay pathways are summarized in Fig. \ref{decay}.

\begin{figure}[]
\includegraphics[scale=0.75]{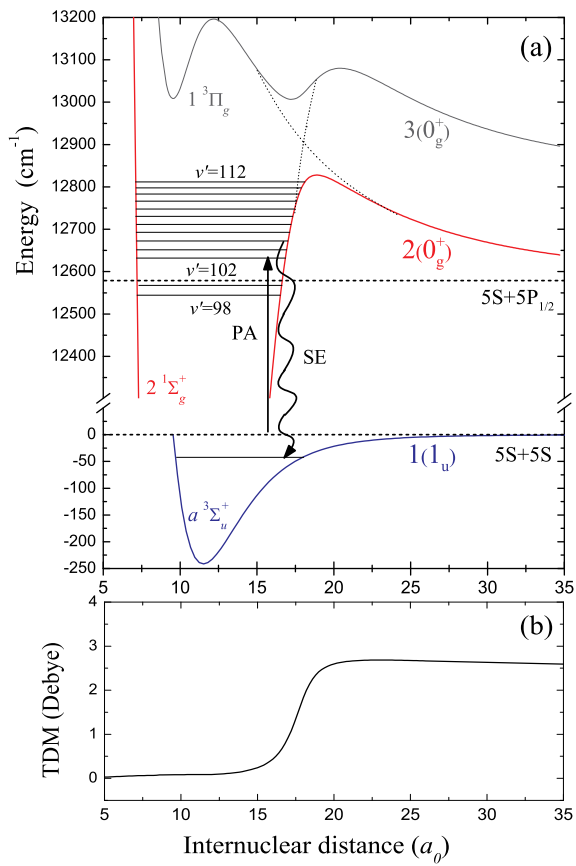}
\caption{\label{pec+tdm}(Color online) Potential energy curves \cite{dulieu, strauss10} showing photoassociation (PA) to 2($0_g^+$) and spontaneous emission (SE) to 1($1_u$). (b) transition dipole moment between the  2($0_g^+$) and 1($1_u$) states, from Ref. \cite{allouche12}.}
\end{figure}

Therefore at short range, in Hund's case (a), levels of the $2\,^1\Sigma_g^+$ state are metastable but at long range, in Hund's case (c), they are not. This is evident in the transition dipole moment (TDM) shown in Fig. \ref{pec+tdm}(b) taken from Allouche and Aubert-Fr\'{e}con \cite{allouche12}.
In the region inside 15 $a_0$, the $2\,^1\Sigma_g^+$ state is well described by Hund's case (a) and the TDM is zero. Levels with both turning points inside 15 $a_0$ should be metastable as the only decay path is through the slow transition to the $A\,^1\Sigma_u^+$ state.

The region between 15 and 20 $a_0$ is where most of the photoassociation and decay occurs in this experiment. The TDM connecting the 2($0_g^+$) and 1($1_u$) states accounts for both the photoassociation step from free triplet colliding atoms and the decay back to 1($1_u$). This region is best described as an intermediate between Hund's case (a) and Hund's case (c) couplings, making all of the allowed Hund's case (a) and Hund's (c) decays shown in Fig. \ref{decay} possible. For example, the branching ratio for decay from $v'=101$ to the $A\,^1\Sigma_u^+$ state compared to the 1($1_u$) state is calculated to be 1:10 based on the ratio of Einstein $A$ coefficients as shown in Table \ref{energies}. The rapid transition to Hund's case (c) between 15 and 20 $a_0$ is due in part to an avoided crossing between the 2($0_g^+$) and 3($0_g^+$) states at 18 $a_0$, as shown in Fig. \ref{pec+tdm}(a), which increases the triplet character of the 2($0_g^+$) state.

The change in coupling to Hund's case (c) occurs roughly between 20 and 40 $a_0$ for most other electronic states in Rb$_2$ \cite{allouche12}. Generally speaking, the range at which this change in coupling occurs is inversely proportional to the strength of the fine structure splitting between the $P_{1/2}$ and $P_{3/2}$ asymptotes. The heavier the alkali dimer, the stronger the fine structure splitting, and therefore the smaller the distance at which the coupling changes from Hund's case (a) to Hund's case (c).

\section{Photoassociation spectroscopy}
\begin{table*}[]
\caption{Energies and assignment ($E_{v', J'}$) of observed levels, with respect to the ground state dissociation limit of two Rb $\left | 5s_{1/2}, F = 3 \right>$ atoms. A single $v'$=112 line is observed at 12811.86 cm$^{-1}$ with an unknown rotational assignment. Systematic uncertainties are $\pm$0.03 cm$^{-1}$, while random uncertainties are $\pm$0.001 cm$^{-1}$. The experimental rotational constants ($B_v$) are given along with fitting uncertainties when known. Also shown are Einstein $A$ coefficients for spontaneous emission from a single vibrational level $v'$ of the $2\,^1\Sigma_g^+$ state to all vibrational levels $v''$ of the $a\,^3\Sigma_u^+ (\Omega=1_u)$ and $A\,^1\Sigma_u^+$ states, denoted by $\sum A_{v' \rightarrow a}$ and $\sum A_{v' \rightarrow A}$ respectively. These $A$ coefficients are calculated using LEVEL 8.0 \cite{level8}. }
\scriptsize
\begin{tabular}{rcccccccccclclcl}
\toprule
  & & & & & $E_{v', J'}$ (cm$^{-1}$) &  &  &  & & & $B_v$ ($10^{-4}$ cm$^{-1}$)& & $\sum A_{v' \rightarrow a}$ ($10^{5}$ s$^{-1}$)& &$\sum A_{v' \rightarrow A}$ ($10^{5}$ s$^{-1}$) \\
\cline{1-9}\cline{11-12}\cline{14-14}\cline{16-16}
$v'$ & & &$J'$=0 &$J'$=1 &$J'$=2&$J'$=3  &$J'$=4  &$J'$=5  & & & && &&  \\

\colrule
98   & &  &  &  &  &12544.882  &  &12545.072& &  & 105.56 & &1.16&&0.177\\
99   & & &  & 12566.818 &12566.860  &12566.953  &12567.006  &12567.112& & & 105.35$\,\pm$0.1& &1.33&&0.175\\
100  & & &  &  &  &   &  & & & & & &1.54&&0.173\\
101  & & &  &  &  &   &  & & & & & &1.80&&0.171\\
102  & & &  &  &  & 12631.135  &  & 12631.321& & &103.22&&2.10&&0.169\\
103  & & &  &  &  & 12651.847 &  &12652.034& & &104.17&&2.45&&0.167\\
104  & & &  &  &  & 12672.143&  & 12672.328& & &102.89&&2.89&&0.164\\
105  & & &  &12691.908  &  & 12692.007  &  &12692.186& & &99.18$\,\pm$0.1 & &3.51&&0.162 \\
106  & & &  &  &  &12711.391  &  &12711.565& & &97.28& &4.35&&0.159\\
107  & & &  &12730.150  &  &12730.236  &  &12730.409& & & 95.97$\,\pm$0.2& &5.41&&0.156\\
108  & & &12748.348  &12748.367  &12748.404  &12748.460  &  &12748.631 & & &94.41$\,\pm$0.3& &6.52&&0.154\\
109  & & & 12765.850 & 12765.869 &12765.905  &12765.960  &  &12766.124 & & &91.25$\,\pm$0.1& &8.80&&0.151\\
110  & & &12782.476  &12782.493  &12782.529  &12782.582  &  &12782.738 & & &87.94$\,\pm$0.1& &12.0&&0.148\\
111  &&  &12797.945  &12797.962  &12797.996  &12798.047  &  &12798.199 & & &84.59$\,\pm$0.1& &17.7&&0.145\\
\botrule
\end{tabular}
\label{energies}
\end{table*}

\begin{figure}[]
\includegraphics[scale=0.75]{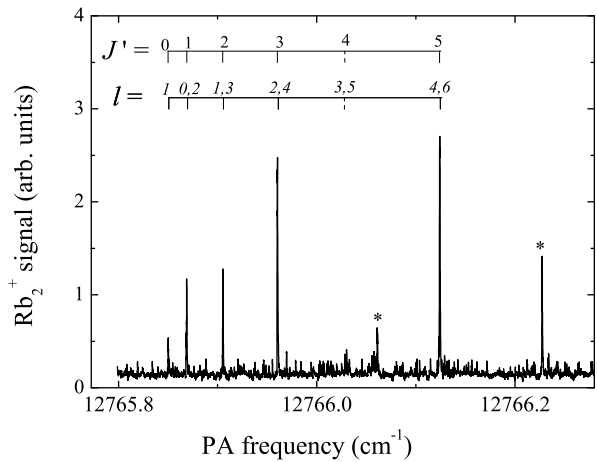}
\caption{\label{v109} PA spectrum of the $2\,^1\Sigma_g^+$ $v'$=109 level showing the rotational assignment ($J'$) and the possible partial wave ($l$) content of each rotational line. The line intensities reveal the presence of partial waves $l=0, 1, 2, 4$ and exclude the presence of partial waves $l=3, 5, 6$. Lines marked with an asterisk (*) are hyperfine ghost lines of the $J'$=3 and 5 transitions as described in the text.}
\end{figure}

The energies of the observed rovibrational levels are listed in Table \ref{energies} and the spectrum of a single vibrational level is shown in Fig. \ref{v109}. The vibrational assignments are determined by comparing the measured vibrational energy spacings ($\Delta$ $G_{v+1/2}$) with those generated from \textit{ab-initio} potential energy curves from Dulieu and Gerdes (DG) \cite{dulieu} and  Allouche and Aubert-Fr\'{e}con (AA) \cite{allouche12}. The DG potential has $v'$=113 as the uppermost vibrational level, while the AA potential has $v'$=126. The experimental vibrational spacings match the DG potential very closely, so we adopt the corresponding vibrational numbering. Vibrational spacings from the bottom of the potential well \cite{amiot87}, on the other hand, match the AA potential somewhat closer.  The rotational assignments are verified by fitting the energies to $B_v[J(J+1)]$, which also determines the rotational constants $B_v$ listed in Table \ref{energies}. We do not take into account small frequency shifts induced by the PA \cite{bohn99,* prodan03} and trapping lasers, which can shift the line position, typically by about 10 MHz. The rotational constants calculated for the DG potentials are larger than the measured rotational constants, while those calculated for the AA potential are smaller.

Levels above $v'$=99 are quasibound and can tunnel through the potential barrier and dissociate into $5s_{1/2}$ and $5p_{1/2}$ atoms. The tunneling probability competes with spontaneous emission only for the two highest vibrational levels. For $v'$=112, the rates are comparable, which decreases the observed signal size. The uppermost vibrational level, $v'$=113, is calculated to dissociate rapidly. Its predicted linewidth (20 GHz) is orders of magnitude broader than for lower vibrational levels, making it very difficult to observe experimentally.

The first two quasibound levels above the atomic limit, $v'=100$ and $v'=101$, are not observed. One possibility is that the absence of these vibrational levels is due to a small FCF overlap for the PA transitions. However, the calculated FCFs for these levels are predicted to be higher than for most of the observed levels. Another possibility is that the PA laser coincidentally couples molecules to higher-lying states, through bound-to-bound or bound-to-free transitions, pumping them away from the detection pathway. Although we cannot rule out accidental bound-bound transitions, it is unlikely for such coincidences to occur for all rotational levels of two successive vibrational levels. Bound-free transitions to a repulsive potential, on the other hand, may occur for a wide range of laser frequencies, but are limited by the wavefunction overlap. This makes it unlikely that bound-free transitions would dominate over the spontaneous decay from the $2\,^1\Sigma_g^+$ state to the $a\,^3\Sigma_u^+$ state. Another explanation is that optical shielding \cite{Sanchez-Villicana95,*suominen95} from the blue-detuned photoassociation beam does not allow atom pairs to reach small interatomic distances and photoassociate. Instead, atom pairs are excited to the outer region of the $2\,^1\Sigma_g^+$ barrier before they can photoassociate at short range to levels inside the barrier. The optical shielding effect is strongest for levels just above the asymptote, namely for $v'=100$, and decreases with energy above the asymptote. Modeling this optical shielding at short internuclear distances is beyond the scope of this work, but offers an avenue for future work. For example, one empirical test for the optical shielding is to monitor the PA signal of a given line and look for reductions in PA signal upon the introduction of an additional blue-detuned laser tuned below the PA line and above the atomic limit. Another test is to search for differences between the intensity dependence of red-detuned photoassociation \cite{bohn99,kraft05} and the intensity dependence of blue-detuned photoassociation.

Other than these missing vibrational levels and the potential for optical shielding, we were unable to find differences between photoassociation to red-detuned and blue-detuned levels. We suspect that effects due to the trapping potentials of red-detuned PA beams---or the anti-trapping potentials of blue-detuned beams---should appear at higher intensities and smaller detunings than those used in the experiment.

We did not search for levels below $v'$=98. We expect their signal size to decrease due to a decreasing TDM as discussed in Sec. III.

\subsection{Rotational levels, partial waves, and shape resonances}

\begin{table*}
\caption{Calculated and previously determined values for the lowest shape resonance in the ground and lowest triplet states of $^{85}$Rb$_2$ and $^{87}$Rb$_2$. The calculated values are obtained using the potentials of Ref. \cite{strauss10}.}
\scriptsize
\begin{tabular}{ccccccccccccc}
\toprule
Molecule&&State&&&Calculated&& && &Previous work&&\\
\cline{1-1}\cline{3-3}\cline{5-7}\cline{9-13}
&&& &  Partial wave,  & Resonance  & Tunneling &$\:$&&  Partial wave,  & Resonance  & Tunneling&  Ref.\\
&& &    & $l$                  & energy  &  width/2$\pi$ &&& $l$                   & energy  &  width/2$\pi$ & \\
&& &&  & (mK) &  (MHz) &&& & (mK)&(MHz) & \\
\colrule
$^{85}$Rb$_2$&& $a\,^3\Sigma_u^+$& & 4&0.66&0.1&&&4& 0.6-0.8 & 0.04-0.16  &\cite{boesten96}\\
$^{85}$Rb$_2$& &$X\,^1\Sigma_g^+$&  &4&0.28&0.002&&&&&&\\
$^{87}$Rb$_2$& &$a\,^3\Sigma_u^+$& &2&0.25&4.5&&& 2 &  & 1.5-7.5 & \cite{boesten97}\\
&& & &&&&&  &2& 0.28 &  & \cite{simoni02}\\
&& & &&&&&  &2& 0.312(25) & 2.4(0.5)  & \cite{voltz05}\\
&& & &&&&&  &2& 0.300(70) & 3 & \cite{buggle04}\\
$^{87}$Rb$_2$& & $X\,^1\Sigma_g^+$& &2&0.33&5.7&&&&&&\\
\botrule
\end{tabular}
\label{shaperesonances}
\end{table*}The distribution of rotational lines of a single vibrational level carries information about the partial wave content of the colliding ground-state atoms. The relative strength of rotational lines in the spectrum is related to the relative partial wave amplitudes. A single rotational line, $J$, is typically made up of several partial waves.

From conservation of angular momentum \cite{jones99,jones06}, we know that $\vec{J}=\vec{l}+\vec{j}$, where $\vec{J}$ is the total angular momentum of the molecule, $\vec{l}$ is the partial wave or ``mechanical rotational quantum'' of the ground-state collision, and $\vec{j}=\vec{j_a}+\vec{j_b}$ is the total atomic electronic angular momentum of both atoms at their asymptotic limit. For a potential curve converging to the $5s_{1/2}+5p_{1/2}$ asymptote; $\vec{j_a}=1/2$ and $\vec{j_b}=1/2$, implying that $j=0,1$. Therefore $J$ can take values $J=l-1$, $l$, or $l+1$. Furthermore, a symmetry consideration \cite{tiesenga96,jones06} requires that in states with (+) symmetry; odd $J$'s come from even $l$'s, and even $J$'s come from odd $l$'s. This additional requirement restricts the values of $J$ to $J=l\pm1$. Therefore, for the $2\,^1\Sigma_g^+$ state, $s$-wave collisions contribute only to the $J'$=1 line, $p$-wave collisions contribute to the $J'$=0 and $J'$=2 lines, and so forth. The $g$-wave collisions---the highest observed partial wave---contribute to the $J'$=3 and $J'$=5 lines, as shown in Fig. \ref{v109}.

\begin{figure}[]
\includegraphics[scale=0.55]{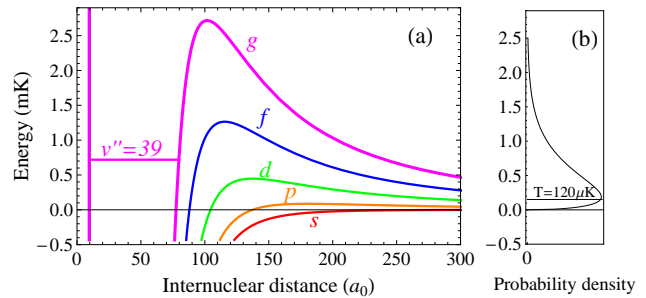}
\caption{\label{partialwavefig} (Color online) (a) Close-up of the $a\,^3\Sigma_u^+$ effective potential curves of $^{85}$Rb$_2$ around the region of the centrifugal barriers for $s$, $p$, $d$, $f$, and $g$-partial waves. The $g$-wave quasibound level responsible for the shape resonance is also shown. (b) The Maxwell-Boltzmann energy distribution of a MOT at 120 $\mu$K showing the high energy tail extending to, and past, the quasibound level.}
\end{figure}

In our spectra, the strongest---and sometimes only---lines are the $J'$=3 and $J'$=5 lines. This implies that the $l$=4 partial wave is the strongest contribution to the PA transitions, clearly at odds with the common notion that $s$-wave scattering dominates processes at ultracold temperatures. This enhancement of the $g$-wave is caused by a shape resonance in the scattering of ground-state atom pairs. This shape resonance is due to the presence of a quasibound level inside the centrifugal potential barrier associated with non-zero angular momentum scattering as is shown in Fig. \ref{partialwavefig}(a). This quasibound level enhances the continuum wavefunction amplitude inside the centrifugal barrier \cite{boesten99,*londono10}. The population of this level depends strongly on the temperature of the system. At temperatures in the quantum degenerate regime (e.g., 1 $\mu$K), we would not expect any significant population of the quasibound level and the shape resonance should not be observable.
The curves in Fig. \ref{partialwavefig}(a) are generated by adding a centrifugal term to the highly accurate $a\,^3\Sigma_u^+$ potential from Strauss $et$ $al$. \cite{strauss10}. The $l$=4 quasibound level was found by numerically solving for bound states with \verb"LEVEL 8.0" \cite{level8}. This rovibrational level ($v''$=39, $l$=4) has a calculated resonance energy of $E$=+0.66 mK and a tunneling width of $\Gamma/2\pi$=0.1 MHz. This same triplet quasibound level has previously been observed for $^{85}$Rb$_2$ by Boesten $et$ $al$. \cite{boesten96} and a corresponding resonance has been observed in $^{87}$Rb$_2$ \cite{boesten97,boesten99,buggle04,thomas04,voltz05}. Shape resonances have also been observed in other alkali dimers, for example, in K$_2$ \cite{demarco99,burke99}, Li$_2$ \cite{cote99}, and NaCs \cite{zabawa11}. In Fig. \ref{partialwavefig}(b) we plot the Maxwell-Boltzmann temperature distribution at the average MOT temperature,

\begin{equation}
f(E)=\sqrt{\frac{E}{\pi k T}}\, \mathrm{exp}\left[{\frac{-E}{kT}}\right],
\end{equation}

showing the overlap of energies between the quasibound state and the energies of the atoms in the MOT. Many of the colliding atom pairs can tunnel through the centrifugal $g$-wave barrier barrier and populate the quasibound state. Since our calculated parameters (energy, width, and partial wave) of this triplet-state shape resonance in $^{85}$Rb$_2$ match experimental values \cite{boesten96} so closely, we have extended the calculation to other combinations of scattering states and isotopologues of Rb$_2$ as shown in Table \ref{shaperesonances}. It is interesting to note that the singlet and triplet states of $^{85}$Rb$_2$ both have $g$-wave shape resonances, albeit with different energies and widths. Similarly, the singlet and triplet states of $^{87}$Rb$_2$ both have a $d$-wave shape resonance. This occurs because the last few vibrational levels of the singlet and triplet states are nearly degenerate \cite{strauss10}, making the singlet and triplet quasibound levels also nearly degenerate.

Unusually high rotational levels, due to mechanisms other than a shape resonance, have been observed by other groups and explained in terms of attraction caused by the trapping laser \cite{fioretti99,*shaffer99}, and attraction caused by dipole trapping from a highly focused PA beam \cite{gomez07}. We do not see evidence for these effects in our work, which as we describe, in this section and the following, is fully accounted for by the shape resonance.

\subsection{Lineshapes}

\begin{figure}[]
\includegraphics[scale=0.75]{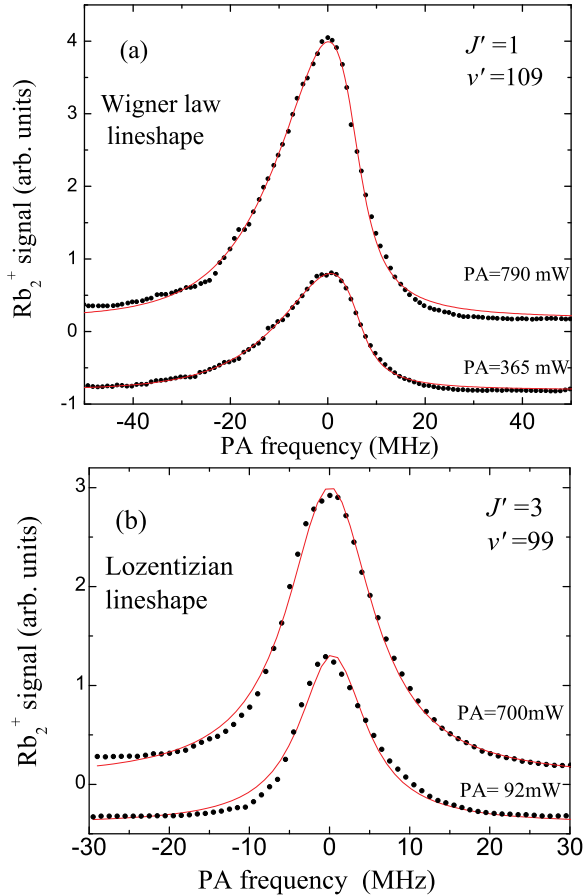}
\caption{\label{lineshapes} (Color online) Experimental ($\bullet$) and fitted (---) lineshapes for $J'$=1 (a) and $J'$=3 (b) at different PA powers. Lines are offset vertically for clarity and horizontally to match the peak position. The small deviation between experimental and fitted lineshape in (b) is not physical and is due to the PA laser scan rate.}
\end{figure}

Another manifestation of shape resonances is through the photoassociation lineshape \cite{simoni02}. In Fig. \ref{lineshapes} we compare the lineshapes arising from $s$-wave and $g$-wave collisions, i.e., the $J'$=1 and the $J'$=3 lines. The $J'$=1 lineshape shown in Fig. \ref{lineshapes} (a) is distinctly asymmetric, with a tail on the red side. The asymmetry is due to the high energy tail in the energy distribution of atoms in the MOT, as shown in Fig. \ref{partialwavefig}(b). These lineshapes were modeled with the ``Wigner law'' lineshape given by Jones $et$ $al.$ \cite{jones99} (Eq.$\,$3). Although the fit is good, the resulting fitting parameters are not physical in the case of Rb$_2$ at a temperature of 120 $\mu$K. This is due to a partial breakdown of the lineshape model, also discussed in Ref. \cite{jones99}. For example, the temperature extracted from the fit is two to three times higher than $120 \pm 20$ $\mu$K measured by ballistic expansion. The lineshapes of $J'$=3 lines are in contrast almost symmetric, implying that only a narrow range of collisional energies participates in the photoassociation process. Furthermore the linewidths for the $J'$=3 lines are generally narrower than for the $J'$=1 lines, again due a narrower range of collisional energies participating in the photoassociation. The ground-state $g$-wave collision in the MOT populates the quasibound level, making the photoassociation more resemble a bound-bound transition than a free-bound transition. Since there is negligible thermal broadening, one can extract the excited state natural linewidths by fitting a simple Lorentizian lineshape.

\section{Conclusion}

We have observed photoassociation to thirteen vibrational levels of the $2\,^1\Sigma_g^+$ state in $^{85}$Rb$_2$. Most of these levels lie above the corresponding $5s_{1/2}+5p_{1/2}$ asymptote. These levels spontaneously decay predominantly to the lowest triplet $a\,^3\Sigma_u^+$ state even though the transition is forbidden by spin selection rules. The presence of a shape resonance causes strong transitions to rotational levels $J'$=3 and 5, and also influences their lineshapes.

\section{ACKNOWLEDGEMENTS}
We thank O. Dulieu and A. Gerdes for providing us with unpublished potential curves, and I. Simbotin for useful discussions. We gratefully acknowledge funding from NSF (PHY-0855613) and AFOSR MURI (FA 9550–-09–-1–-0588)

\bibliography{C:/Users/Public/bibliography/bibtex_masterfile_donotdelete}

\begin{thebibliography}{46}%
\makeatletter
\providecommand \@ifxundefined [1]{%
 \@ifx{#1\undefined}
}%
\providecommand \@ifnum [1]{%
 \ifnum #1\expandafter \@firstoftwo
 \else \expandafter \@secondoftwo
 \fi
}%
\providecommand \@ifx [1]{%
 \ifx #1\expandafter \@firstoftwo
 \else \expandafter \@secondoftwo
 \fi
}%
\providecommand \natexlab [1]{#1}%
\providecommand \enquote  [1]{``#1''}%
\providecommand \bibnamefont  [1]{#1}%
\providecommand \bibfnamefont [1]{#1}%
\providecommand \citenamefont [1]{#1}%
\providecommand \href@noop [0]{\@secondoftwo}%
\providecommand \href [0]{\begingroup \@sanitize@url \@href}%
\providecommand \@href[1]{\@@startlink{#1}\@@href}%
\providecommand \@@href[1]{\endgroup#1\@@endlink}%
\providecommand \@sanitize@url [0]{\catcode `\\12\catcode `\$12\catcode
  `\&12\catcode `\#12\catcode `\^12\catcode `\_12\catcode `\%12\relax}%
\providecommand \@@startlink[1]{}%
\providecommand \@@endlink[0]{}%
\providecommand \url  [0]{\begingroup\@sanitize@url \@url }%
\providecommand \@url [1]{\endgroup\@href {#1}{\urlprefix }}%
\providecommand \urlprefix  [0]{URL }%
\providecommand \Eprint [0]{\href }%
\providecommand \doibase [0]{http://dx.doi.org/}%
\providecommand \selectlanguage [0]{\@gobble}%
\providecommand \bibinfo  [0]{\@secondoftwo}%
\providecommand \bibfield  [0]{\@secondoftwo}%
\providecommand \translation [1]{[#1]}%
\providecommand \BibitemOpen [0]{}%
\providecommand \bibitemStop [0]{}%
\providecommand \bibitemNoStop [0]{.\EOS\space}%
\providecommand \EOS [0]{\spacefactor3000\relax}%
\providecommand \BibitemShut  [1]{\csname bibitem#1\endcsname}%
\let\auto@bib@innerbib\@empty
\bibitem [{\citenamefont {Stwalley}\ and\ \citenamefont
  {Wang}(1999)}]{stwalley99}%
  \BibitemOpen
  \bibfield  {author} {\bibinfo {author} {\bibfnamefont {W.~C.}\ \bibnamefont
  {Stwalley}}\ and\ \bibinfo {author} {\bibfnamefont {H.}~\bibnamefont
  {Wang}},\ }\href {\doibase 10.1006/jmsp.1999.7838} {\bibfield  {journal}
  {\bibinfo  {journal} {J. Mol. Spectrosc.}\ }\textbf {\bibinfo {volume}
  {195}},\ \bibinfo {pages} {194 } (\bibinfo {year} {1999})}\BibitemShut
  {NoStop}%
\bibitem [{\citenamefont {Jones}\ \emph {et~al.}(2006)\citenamefont {Jones},
  \citenamefont {Tiesinga}, \citenamefont {Lett},\ and\ \citenamefont
  {Julienne}}]{jones06}%
  \BibitemOpen
  \bibfield  {author} {\bibinfo {author} {\bibfnamefont {K.~M.}\ \bibnamefont
  {Jones}}, \bibinfo {author} {\bibfnamefont {E.}~\bibnamefont {Tiesinga}},
  \bibinfo {author} {\bibfnamefont {P.~D.}\ \bibnamefont {Lett}}, \ and\
  \bibinfo {author} {\bibfnamefont {P.~S.}\ \bibnamefont {Julienne}},\ }\href
  {\doibase 10.1103/RevModPhys.78.483} {\bibfield  {journal} {\bibinfo
  {journal} {Rev. Mod. Phys.}\ }\textbf {\bibinfo {volume} {78}},\ \bibinfo
  {pages} {483} (\bibinfo {year} {2006})}\BibitemShut {NoStop}%
\bibitem [{\citenamefont {Carr}\ \emph {et~al.}(2009)\citenamefont {Carr},
  \citenamefont {DeMille}, \citenamefont {Krems},\ and\ \citenamefont
  {Ye}}]{carr09}%
  \BibitemOpen
  \bibfield  {author} {\bibinfo {author} {\bibfnamefont {L.~D.}\ \bibnamefont
  {Carr}}, \bibinfo {author} {\bibfnamefont {D.}~\bibnamefont {DeMille}},
  \bibinfo {author} {\bibfnamefont {R.~V.}\ \bibnamefont {Krems}}, \ and\
  \bibinfo {author} {\bibfnamefont {J.}~\bibnamefont {Ye}},\ }\href
  {http://stacks.iop.org/1367-2630/11/i=5/a=055049} {\bibfield  {journal}
  {\bibinfo  {journal} {New J. Phys.}\ }\textbf {\bibinfo {volume} {11}},\
  \bibinfo {pages} {055049} (\bibinfo {year} {2009})}\BibitemShut {NoStop}%
\bibitem [{\citenamefont {Krems}(2008)}]{krems08}%
  \BibitemOpen
  \bibfield  {author} {\bibinfo {author} {\bibfnamefont {R.~V.}\ \bibnamefont
  {Krems}},\ }\href {\doibase 10.1039/B802322K} {\bibfield  {journal} {\bibinfo
   {journal} {Phys. Chem. Chem. Phys.}\ }\textbf {\bibinfo {volume} {10}},\
  \bibinfo {pages} {4079} (\bibinfo {year} {2008})}\BibitemShut {NoStop}%
\bibitem [{\citenamefont {Hutzler}\ \emph {et~al.}(2012)\citenamefont
  {Hutzler}, \citenamefont {Lu},\ and\ \citenamefont {Doyle}}]{hutzler12}%
  \BibitemOpen
  \bibfield  {author} {\bibinfo {author} {\bibfnamefont {N.~R.}\ \bibnamefont
  {Hutzler}}, \bibinfo {author} {\bibfnamefont {H.-I.}\ \bibnamefont {Lu}}, \
  and\ \bibinfo {author} {\bibfnamefont {J.~M.}\ \bibnamefont {Doyle}},\ }\href
  {\doibase DOI: 10.1021/cr200362u} {\bibfield  {journal} {\bibinfo  {journal}
  {Chem. Rev.}\ } (\bibinfo {year} {2012}),\ DOI:
  10.1021/cr200362u}\BibitemShut {NoStop}%
\bibitem [{\citenamefont {Bethlem}\ \emph {et~al.}(1999)\citenamefont
  {Bethlem}, \citenamefont {Berden},\ and\ \citenamefont {Meijer}}]{bethlem99}%
  \BibitemOpen
  \bibfield  {author} {\bibinfo {author} {\bibfnamefont {H.~L.}\ \bibnamefont
  {Bethlem}}, \bibinfo {author} {\bibfnamefont {G.}~\bibnamefont {Berden}}, \
  and\ \bibinfo {author} {\bibfnamefont {G.}~\bibnamefont {Meijer}},\ }\href
  {\doibase 10.1103/PhysRevLett.83.1558} {\bibfield  {journal} {\bibinfo
  {journal} {Phys. Rev. Lett.}\ }\textbf {\bibinfo {volume} {83}},\ \bibinfo
  {pages} {1558} (\bibinfo {year} {1999})}\BibitemShut {NoStop}%
\bibitem [{\citenamefont {Winkler}\ \emph {et~al.}(2007)\citenamefont
  {Winkler}, \citenamefont {Lang}, \citenamefont {Thalhammer}, \citenamefont
  {van~der Straten}, \citenamefont {Grimm},\ and\ \citenamefont
  {Denschlag}}]{winkler07}%
  \BibitemOpen
  \bibfield  {author} {\bibinfo {author} {\bibfnamefont {K.}~\bibnamefont
  {Winkler}}, \bibinfo {author} {\bibfnamefont {F.}~\bibnamefont {Lang}},
  \bibinfo {author} {\bibfnamefont {G.}~\bibnamefont {Thalhammer}}, \bibinfo
  {author} {\bibfnamefont {P.}~\bibnamefont {van~der Straten}}, \bibinfo
  {author} {\bibfnamefont {R.}~\bibnamefont {Grimm}}, \ and\ \bibinfo {author}
  {\bibfnamefont {J.~H.}\ \bibnamefont {Denschlag}},\ }\href {\doibase
  10.1103/PhysRevLett.98.043201} {\bibfield  {journal} {\bibinfo  {journal}
  {Phys. Rev. Lett.}\ }\textbf {\bibinfo {volume} {98}},\ \bibinfo {pages}
  {043201} (\bibinfo {year} {2007})}\BibitemShut {NoStop}%
\bibitem [{\citenamefont {Aikawa}\ \emph {et~al.}(2010)\citenamefont {Aikawa},
  \citenamefont {Akamatsu}, \citenamefont {Hayashi}, \citenamefont {Oasa},
  \citenamefont {Kobayashi}, \citenamefont {Naidon}, \citenamefont {Kishimoto},
  \citenamefont {Ueda},\ and\ \citenamefont {Inouye}}]{aikawa10}%
  \BibitemOpen
  \bibfield  {author} {\bibinfo {author} {\bibfnamefont {K.}~\bibnamefont
  {Aikawa}}, \bibinfo {author} {\bibfnamefont {D.}~\bibnamefont {Akamatsu}},
  \bibinfo {author} {\bibfnamefont {M.}~\bibnamefont {Hayashi}}, \bibinfo
  {author} {\bibfnamefont {K.}~\bibnamefont {Oasa}}, \bibinfo {author}
  {\bibfnamefont {J.}~\bibnamefont {Kobayashi}}, \bibinfo {author}
  {\bibfnamefont {P.}~\bibnamefont {Naidon}}, \bibinfo {author} {\bibfnamefont
  {T.}~\bibnamefont {Kishimoto}}, \bibinfo {author} {\bibfnamefont
  {M.}~\bibnamefont {Ueda}}, \ and\ \bibinfo {author} {\bibfnamefont
  {S.}~\bibnamefont {Inouye}},\ }\href {\doibase
  10.1103/PhysRevLett.105.203001} {\bibfield  {journal} {\bibinfo  {journal}
  {Phys. Rev. Lett.}\ }\textbf {\bibinfo {volume} {105}},\ \bibinfo {pages}
  {203001} (\bibinfo {year} {2010})}\BibitemShut {NoStop}%
\bibitem [{\citenamefont {Barry}\ \emph {et~al.}(2012)\citenamefont {Barry},
  \citenamefont {Shuman}, \citenamefont {Norrgard},\ and\ \citenamefont
  {DeMille}}]{barry12}%
  \BibitemOpen
  \bibfield  {author} {\bibinfo {author} {\bibfnamefont {J.~F.}\ \bibnamefont
  {Barry}}, \bibinfo {author} {\bibfnamefont {E.~S.}\ \bibnamefont {Shuman}},
  \bibinfo {author} {\bibfnamefont {E.~B.}\ \bibnamefont {Norrgard}}, \ and\
  \bibinfo {author} {\bibfnamefont {D.}~\bibnamefont {DeMille}},\ }\href
  {\doibase 10.1103/PhysRevLett.108.103002} {\bibfield  {journal} {\bibinfo
  {journal} {Phys. Rev. Lett.}\ }\textbf {\bibinfo {volume} {108}},\ \bibinfo
  {pages} {103002} (\bibinfo {year} {2012})}\BibitemShut {NoStop}%
\bibitem [{\citenamefont {Gupta}\ and\ \citenamefont
  {Herschbach}(1999)}]{gupta99}%
  \BibitemOpen
  \bibfield  {author} {\bibinfo {author} {\bibfnamefont {M.}~\bibnamefont
  {Gupta}}\ and\ \bibinfo {author} {\bibfnamefont {D.}~\bibnamefont
  {Herschbach}},\ }\href {\doibase 10.1021/jp993560x} {\bibfield  {journal}
  {\bibinfo  {journal} {J. Phys. Chem. A}\ }\textbf {\bibinfo {volume} {103}},\
  \bibinfo {pages} {10670} (\bibinfo {year} {1999})}\BibitemShut {NoStop}%
\bibitem [{\citenamefont {Deiglmayr}\ \emph {et~al.}(2008)\citenamefont
  {Deiglmayr}, \citenamefont {Grochola}, \citenamefont {Repp}, \citenamefont
  {M\"ortlbauer}, \citenamefont {Gl\"uck}, \citenamefont {Lange}, \citenamefont
  {Dulieu}, \citenamefont {Wester},\ and\ \citenamefont
  {Weidem\"uller}}]{deiglmayr08}%
  \BibitemOpen
  \bibfield  {author} {\bibinfo {author} {\bibfnamefont {J.}~\bibnamefont
  {Deiglmayr}}, \bibinfo {author} {\bibfnamefont {A.}~\bibnamefont {Grochola}},
  \bibinfo {author} {\bibfnamefont {M.}~\bibnamefont {Repp}}, \bibinfo {author}
  {\bibfnamefont {K.}~\bibnamefont {M\"ortlbauer}}, \bibinfo {author}
  {\bibfnamefont {C.}~\bibnamefont {Gl\"uck}}, \bibinfo {author} {\bibfnamefont
  {J.}~\bibnamefont {Lange}}, \bibinfo {author} {\bibfnamefont
  {O.}~\bibnamefont {Dulieu}}, \bibinfo {author} {\bibfnamefont
  {R.}~\bibnamefont {Wester}}, \ and\ \bibinfo {author} {\bibfnamefont
  {M.}~\bibnamefont {Weidem\"uller}},\ }\href {\doibase
  10.1103/PhysRevLett.101.133004} {\bibfield  {journal} {\bibinfo  {journal}
  {Phys. Rev. Lett.}\ }\textbf {\bibinfo {volume} {101}},\ \bibinfo {pages}
  {133004} (\bibinfo {year} {2008})}\BibitemShut {NoStop}%
\bibitem [{\citenamefont {Bellos}\ \emph {et~al.}(2011)\citenamefont {Bellos},
  \citenamefont {Rahmlow}, \citenamefont {Carollo}, \citenamefont {Banerjee},
  \citenamefont {Dulieu}, \citenamefont {Gerdes}, \citenamefont {Eyler},
  \citenamefont {Gould},\ and\ \citenamefont {Stwalley}}]{bellos11}%
  \BibitemOpen
  \bibfield  {author} {\bibinfo {author} {\bibfnamefont {M.~A.}\ \bibnamefont
  {Bellos}}, \bibinfo {author} {\bibfnamefont {D.}~\bibnamefont {Rahmlow}},
  \bibinfo {author} {\bibfnamefont {R.}~\bibnamefont {Carollo}}, \bibinfo
  {author} {\bibfnamefont {J.}~\bibnamefont {Banerjee}}, \bibinfo {author}
  {\bibfnamefont {O.}~\bibnamefont {Dulieu}}, \bibinfo {author} {\bibfnamefont
  {A.}~\bibnamefont {Gerdes}}, \bibinfo {author} {\bibfnamefont {E.~E.}\
  \bibnamefont {Eyler}}, \bibinfo {author} {\bibfnamefont {P.~L.}\ \bibnamefont
  {Gould}}, \ and\ \bibinfo {author} {\bibfnamefont {W.~C.}\ \bibnamefont
  {Stwalley}},\ }\href {\doibase 10.1039/C1CP21383K} {\bibfield  {journal}
  {\bibinfo  {journal} {Phys. Chem. Chem. Phys.}\ }\textbf {\bibinfo {volume}
  {13}},\ \bibinfo {pages} {18880} (\bibinfo {year} {2011})}\BibitemShut
  {NoStop}%
\bibitem [{\citenamefont {Gabbanini}\ and\ \citenamefont
  {Dulieu}(2011)}]{gabbanini11}%
  \BibitemOpen
  \bibfield  {author} {\bibinfo {author} {\bibfnamefont {C.}~\bibnamefont
  {Gabbanini}}\ and\ \bibinfo {author} {\bibfnamefont {O.}~\bibnamefont
  {Dulieu}},\ }\href {\doibase 10.1039/C1CP21497G} {\bibfield  {journal}
  {\bibinfo  {journal} {Phys. Chem. Chem. Phys.}\ }\textbf {\bibinfo {volume}
  {13}},\ \bibinfo {pages} {18905} (\bibinfo {year} {2011})}\BibitemShut
  {NoStop}%
\bibitem [{\citenamefont {Ji}\ \emph {et~al.}(2012)\citenamefont {Ji},
  \citenamefont {Zhang}, \citenamefont {Wu}, \citenamefont {Yuan},
  \citenamefont {Yang}, \citenamefont {Zhao}, \citenamefont {Ma}, \citenamefont
  {Wang}, \citenamefont {Xiao},\ and\ \citenamefont {Jia}}]{rbcs2012}%
  \BibitemOpen
  \bibfield  {author} {\bibinfo {author} {\bibfnamefont {Z.}~\bibnamefont
  {Ji}}, \bibinfo {author} {\bibfnamefont {H.}~\bibnamefont {Zhang}}, \bibinfo
  {author} {\bibfnamefont {J.}~\bibnamefont {Wu}}, \bibinfo {author}
  {\bibfnamefont {J.}~\bibnamefont {Yuan}}, \bibinfo {author} {\bibfnamefont
  {Y.}~\bibnamefont {Yang}}, \bibinfo {author} {\bibfnamefont {Y.}~\bibnamefont
  {Zhao}}, \bibinfo {author} {\bibfnamefont {J.}~\bibnamefont {Ma}}, \bibinfo
  {author} {\bibfnamefont {L.}~\bibnamefont {Wang}}, \bibinfo {author}
  {\bibfnamefont {L.}~\bibnamefont {Xiao}}, \ and\ \bibinfo {author}
  {\bibfnamefont {S.}~\bibnamefont {Jia}},\ }\href {\doibase
  10.1103/PhysRevA.85.013401} {\bibfield  {journal} {\bibinfo  {journal} {Phys.
  Rev. A}\ }\textbf {\bibinfo {volume} {85}},\ \bibinfo {pages} {013401}
  (\bibinfo {year} {2012})}\BibitemShut {NoStop}%
\bibitem [{\citenamefont {Zabawa}\ \emph {et~al.}(2011)\citenamefont {Zabawa},
  \citenamefont {Wakim}, \citenamefont {Haruza},\ and\ \citenamefont
  {Bigelow}}]{zabawa11}%
  \BibitemOpen
  \bibfield  {author} {\bibinfo {author} {\bibfnamefont {P.}~\bibnamefont
  {Zabawa}}, \bibinfo {author} {\bibfnamefont {A.}~\bibnamefont {Wakim}},
  \bibinfo {author} {\bibfnamefont {M.}~\bibnamefont {Haruza}}, \ and\ \bibinfo
  {author} {\bibfnamefont {N.~P.}\ \bibnamefont {Bigelow}},\ }\href {\doibase
  10.1103/PhysRevA.84.061401} {\bibfield  {journal} {\bibinfo  {journal} {Phys.
  Rev. A}\ }\textbf {\bibinfo {volume} {84}},\ \bibinfo {pages} {061401}
  (\bibinfo {year} {2011})}\BibitemShut {NoStop}%
\bibitem [{\citenamefont {Chin}\ \emph {et~al.}(2010)\citenamefont {Chin},
  \citenamefont {Grimm}, \citenamefont {Julienne},\ and\ \citenamefont
  {Tiesinga}}]{chin10}%
  \BibitemOpen
  \bibfield  {author} {\bibinfo {author} {\bibfnamefont {C.}~\bibnamefont
  {Chin}}, \bibinfo {author} {\bibfnamefont {R.}~\bibnamefont {Grimm}},
  \bibinfo {author} {\bibfnamefont {P.}~\bibnamefont {Julienne}}, \ and\
  \bibinfo {author} {\bibfnamefont {E.}~\bibnamefont {Tiesinga}},\ }\href
  {\doibase 10.1103/RevModPhys.82.1225} {\bibfield  {journal} {\bibinfo
  {journal} {Rev. Mod. Phys.}\ }\textbf {\bibinfo {volume} {82}},\ \bibinfo
  {pages} {1225} (\bibinfo {year} {2010})}\BibitemShut {NoStop}%
\bibitem [{\citenamefont {Lozeille}\ \emph {et~al.}(2006)\citenamefont
  {Lozeille}, \citenamefont {Fioretti}, \citenamefont {Gabbanini},
  \citenamefont {Huang}, \citenamefont {Pechkis}, \citenamefont {Wang},
  \citenamefont {Gould}, \citenamefont {Eyler}, \citenamefont {Stwalley},
  \citenamefont {Aymar},\ and\ \citenamefont {Dulieu}}]{lozeille06}%
  \BibitemOpen
  \bibfield  {author} {\bibinfo {author} {\bibfnamefont {J.}~\bibnamefont
  {Lozeille}}, \bibinfo {author} {\bibfnamefont {A.}~\bibnamefont {Fioretti}},
  \bibinfo {author} {\bibfnamefont {C.}~\bibnamefont {Gabbanini}}, \bibinfo
  {author} {\bibfnamefont {Y.}~\bibnamefont {Huang}}, \bibinfo {author}
  {\bibfnamefont {H.~K.}\ \bibnamefont {Pechkis}}, \bibinfo {author}
  {\bibfnamefont {D.}~\bibnamefont {Wang}}, \bibinfo {author} {\bibfnamefont
  {P.~L.}\ \bibnamefont {Gould}}, \bibinfo {author} {\bibfnamefont {E.~E.}\
  \bibnamefont {Eyler}}, \bibinfo {author} {\bibfnamefont {W.~C.}\ \bibnamefont
  {Stwalley}}, \bibinfo {author} {\bibfnamefont {M.}~\bibnamefont {Aymar}}, \
  and\ \bibinfo {author} {\bibfnamefont {O.}~\bibnamefont {Dulieu}},\ }\href
  {http://dx.doi.org/10.1140/epjd/e2006-00084-4} {\bibfield  {journal}
  {\bibinfo  {journal} {Eur. Phys. J. D}\ }\textbf {\bibinfo {volume} {39}},\
  \bibinfo {pages} {261} (\bibinfo {year} {2006})}\BibitemShut {NoStop}%
\bibitem [{\citenamefont {Boesten}\ \emph {et~al.}(1996)\citenamefont
  {Boesten}, \citenamefont {Tsai}, \citenamefont {Verhaar},\ and\ \citenamefont
  {Heinzen}}]{boesten96}%
  \BibitemOpen
  \bibfield  {author} {\bibinfo {author} {\bibfnamefont {H.~M. J.~M.}\
  \bibnamefont {Boesten}}, \bibinfo {author} {\bibfnamefont {C.~C.}\
  \bibnamefont {Tsai}}, \bibinfo {author} {\bibfnamefont {B.~J.}\ \bibnamefont
  {Verhaar}}, \ and\ \bibinfo {author} {\bibfnamefont {D.~J.}\ \bibnamefont
  {Heinzen}},\ }\href {\doibase 10.1103/PhysRevLett.77.5194} {\bibfield
  {journal} {\bibinfo  {journal} {Phys. Rev. Lett.}\ }\textbf {\bibinfo
  {volume} {77}},\ \bibinfo {pages} {5194} (\bibinfo {year}
  {1996})}\BibitemShut {NoStop}%
\bibitem [{\citenamefont {Boesten}\ \emph {et~al.}(1999)\citenamefont
  {Boesten}, \citenamefont {Tsai}, \citenamefont {Heinzen}, \citenamefont
  {Moonen},\ and\ \citenamefont {Verhaar}}]{boesten99}%
  \BibitemOpen
  \bibfield  {author} {\bibinfo {author} {\bibfnamefont {H.~M. J.~M.}\
  \bibnamefont {Boesten}}, \bibinfo {author} {\bibfnamefont {C.~C.}\
  \bibnamefont {Tsai}}, \bibinfo {author} {\bibfnamefont {D.~J.}\ \bibnamefont
  {Heinzen}}, \bibinfo {author} {\bibfnamefont {A.~J.}\ \bibnamefont {Moonen}},
  \ and\ \bibinfo {author} {\bibfnamefont {B.~J.}\ \bibnamefont {Verhaar}},\
  }\href {http://stacks.iop.org/0953-4075/32/i=2/a=012} {\bibfield  {journal}
  {\bibinfo  {journal} {J. Phys. B}\ }\textbf {\bibinfo {volume} {32}},\
  \bibinfo {pages} {287} (\bibinfo {year} {1999})}\BibitemShut {NoStop}%
\bibitem [{\citenamefont {Amiot}\ and\ \citenamefont {Verges}(1987)}]{amiot87}%
  \BibitemOpen
  \bibfield  {author} {\bibinfo {author} {\bibfnamefont {C.}~\bibnamefont
  {Amiot}}\ and\ \bibinfo {author} {\bibfnamefont {J.}~\bibnamefont {Verges}},\
  }\href@noop {} {\bibfield  {journal} {\bibinfo  {journal} {Mol. Phys.}\
  }\textbf {\bibinfo {volume} {61}},\ \bibinfo {pages} {51} (\bibinfo {year}
  {1987})}\BibitemShut {NoStop}%
\bibitem [{\citenamefont {Allouche}\ and\ \citenamefont
  {Aubert-Fr\'{e}con}(2012)}]{allouche12}%
  \BibitemOpen
  \bibfield  {author} {\bibinfo {author} {\bibfnamefont {A.-R.}\ \bibnamefont
  {Allouche}}\ and\ \bibinfo {author} {\bibfnamefont {M.}~\bibnamefont
  {Aubert-Fr\'{e}con}},\ }\href {\doibase 10.1063/1.3694014} {\bibfield
  {journal} {\bibinfo  {journal} {J. Chem. Phys.}\ }\textbf {\bibinfo {volume}
  {136}},\ \bibinfo {eid} {114302} (\bibinfo {year} {2012})}\BibitemShut
  {NoStop}%
\bibitem [{\citenamefont {Jraij}\ \emph {et~al.}(2003)\citenamefont {Jraij},
  \citenamefont {Allouche}, \citenamefont {Korek},\ and\ \citenamefont
  {Aubert-Fr\'{e}con}}]{jraij03}%
  \BibitemOpen
  \bibfield  {author} {\bibinfo {author} {\bibfnamefont {A.}~\bibnamefont
  {Jraij}}, \bibinfo {author} {\bibfnamefont {A.}~\bibnamefont {Allouche}},
  \bibinfo {author} {\bibfnamefont {M.}~\bibnamefont {Korek}}, \ and\ \bibinfo
  {author} {\bibfnamefont {M.}~\bibnamefont {Aubert-Fr\'{e}con}},\ }\href
  {\doibase 10.1016/S0301-0104(03)00060-0} {\bibfield  {journal} {\bibinfo
  {journal} {Chem. Phys.}\ }\textbf {\bibinfo {volume} {290}},\ \bibinfo
  {pages} {129 } (\bibinfo {year} {2003})}\BibitemShut {NoStop}%
\bibitem [{\citenamefont {Stwalley}\ \emph {et~al.}(2012)\citenamefont
  {Stwalley}, \citenamefont {Bellos}, \citenamefont {Carollo}, \citenamefont
  {Banerjee},\ and\ \citenamefont {Bermudez}}]{stwalley12}%
  \BibitemOpen
  \bibfield  {author} {\bibinfo {author} {\bibfnamefont {W.~C.}\ \bibnamefont
  {Stwalley}}, \bibinfo {author} {\bibfnamefont {M.}~\bibnamefont {Bellos}},
  \bibinfo {author} {\bibfnamefont {R.}~\bibnamefont {Carollo}}, \bibinfo
  {author} {\bibfnamefont {J.}~\bibnamefont {Banerjee}}, \ and\ \bibinfo
  {author} {\bibfnamefont {M.}~\bibnamefont {Bermudez}},\ }\href {\doibase
  10.1080/00268976.2012.676680} {\bibfield  {journal} {\bibinfo  {journal}
  {Molecular Physics}\ }\textbf {\bibinfo {volume} {110}},\ \bibinfo {pages}
  {1739} (\bibinfo {year} {2012})}\BibitemShut {NoStop}%
\bibitem [{\citenamefont {Viteau}\ \emph {et~al.}(2009)\citenamefont {Viteau},
  \citenamefont {Chotia}, \citenamefont {Allegrini}, \citenamefont {Bouloufa},
  \citenamefont {Dulieu}, \citenamefont {Comparat},\ and\ \citenamefont
  {Pillet}}]{viteau09}%
  \BibitemOpen
  \bibfield  {author} {\bibinfo {author} {\bibfnamefont {M.}~\bibnamefont
  {Viteau}}, \bibinfo {author} {\bibfnamefont {A.}~\bibnamefont {Chotia}},
  \bibinfo {author} {\bibfnamefont {M.}~\bibnamefont {Allegrini}}, \bibinfo
  {author} {\bibfnamefont {N.}~\bibnamefont {Bouloufa}}, \bibinfo {author}
  {\bibfnamefont {O.}~\bibnamefont {Dulieu}}, \bibinfo {author} {\bibfnamefont
  {D.}~\bibnamefont {Comparat}}, \ and\ \bibinfo {author} {\bibfnamefont
  {P.}~\bibnamefont {Pillet}},\ }\href {\doibase 10.1103/PhysRevA.79.021402}
  {\bibfield  {journal} {\bibinfo  {journal} {Phys. Rev. A}\ }\textbf {\bibinfo
  {volume} {79}},\ \bibinfo {pages} {021402} (\bibinfo {year}
  {2009})}\BibitemShut {NoStop}%
\bibitem [{\citenamefont {Dulieu}\ and\ \citenamefont {Gerdes}()}]{dulieu}%
  \BibitemOpen
  \bibfield  {author} {\bibinfo {author} {\bibfnamefont {O.}~\bibnamefont
  {Dulieu}}\ and\ \bibinfo {author} {\bibfnamefont {A.}~\bibnamefont
  {Gerdes}},\ }\href@noop {} {}\bibinfo {note} {Private communication
  (2011)}\BibitemShut {NoStop}%
\bibitem [{\citenamefont {Strauss}\ \emph {et~al.}(2010)\citenamefont
  {Strauss}, \citenamefont {Takekoshi}, \citenamefont {Lang}, \citenamefont
  {Winkler}, \citenamefont {Grimm}, \citenamefont {Hecker~Denschlag},\ and\
  \citenamefont {Tiemann}}]{strauss10}%
  \BibitemOpen
  \bibfield  {author} {\bibinfo {author} {\bibfnamefont {C.}~\bibnamefont
  {Strauss}}, \bibinfo {author} {\bibfnamefont {T.}~\bibnamefont {Takekoshi}},
  \bibinfo {author} {\bibfnamefont {F.}~\bibnamefont {Lang}}, \bibinfo {author}
  {\bibfnamefont {K.}~\bibnamefont {Winkler}}, \bibinfo {author} {\bibfnamefont
  {R.}~\bibnamefont {Grimm}}, \bibinfo {author} {\bibfnamefont
  {J.}~\bibnamefont {Hecker~Denschlag}}, \ and\ \bibinfo {author}
  {\bibfnamefont {E.}~\bibnamefont {Tiemann}},\ }\href {\doibase
  10.1103/PhysRevA.82.052514} {\bibfield  {journal} {\bibinfo  {journal} {Phys.
  Rev. A}\ }\textbf {\bibinfo {volume} {82}},\ \bibinfo {pages} {052514}
  (\bibinfo {year} {2010})}\BibitemShut {NoStop}%
\bibitem [{\citenamefont {Le{$\:$}Roy}(2007)}]{level8}%
  \BibitemOpen
  \bibfield  {author} {\bibinfo {author} {\bibfnamefont {R.~J.}\ \bibnamefont
  {Le{$\:$}Roy}},\ }\href {http://scienide2.uwaterloo.ca/~rleroy/level/}
  {\bibfield  {journal} {\bibinfo  {journal} {LEVEL 8.0: A Computer Program for
  Solving the Radial Schr{\"{o}}odinger Equation for Bound and Quasibound
  Levels, University of Waterloo Chemical Physics Research Report CP-663}\ }
  (\bibinfo {year} {2007})}\BibitemShut {NoStop}%
\bibitem [{\citenamefont {Bohn}\ and\ \citenamefont {Julienne}(1999)}]{bohn99}%
  \BibitemOpen
  \bibfield  {author} {\bibinfo {author} {\bibfnamefont {J.~L.}\ \bibnamefont
  {Bohn}}\ and\ \bibinfo {author} {\bibfnamefont {P.~S.}\ \bibnamefont
  {Julienne}},\ }\href {\doibase 10.1103/PhysRevA.60.414} {\bibfield  {journal}
  {\bibinfo  {journal} {Phys. Rev. A}\ }\textbf {\bibinfo {volume} {60}},\
  \bibinfo {pages} {414} (\bibinfo {year} {1999})}\BibitemShut {NoStop}%
\bibitem [{\citenamefont {Prodan}\ \emph {et~al.}(2003)\citenamefont {Prodan},
  \citenamefont {Pichler}, \citenamefont {Junker}, \citenamefont {Hulet},\ and\
  \citenamefont {Bohn}}]{prodan03}%
  \BibitemOpen
  \bibfield  {author} {\bibinfo {author} {\bibfnamefont {I.~D.}\ \bibnamefont
  {Prodan}}, \bibinfo {author} {\bibfnamefont {M.}~\bibnamefont {Pichler}},
  \bibinfo {author} {\bibfnamefont {M.}~\bibnamefont {Junker}}, \bibinfo
  {author} {\bibfnamefont {R.~G.}\ \bibnamefont {Hulet}}, \ and\ \bibinfo
  {author} {\bibfnamefont {J.~L.}\ \bibnamefont {Bohn}},\ }\href {\doibase
  10.1103/PhysRevLett.91.080402} {\bibfield  {journal} {\bibinfo  {journal}
  {Phys. Rev. Lett.}\ }\textbf {\bibinfo {volume} {91}},\ \bibinfo {pages}
  {080402} (\bibinfo {year} {2003})}\BibitemShut {NoStop}%
\bibitem [{\citenamefont {Sanchez-Villicana}\ \emph {et~al.}(1995)\citenamefont
  {Sanchez-Villicana}, \citenamefont {Gensemer}, \citenamefont {Tan},
  \citenamefont {Kumarakrishnan}, \citenamefont {Dinneen}, \citenamefont
  {S\"uptitz},\ and\ \citenamefont {Gould}}]{Sanchez-Villicana95}%
  \BibitemOpen
  \bibfield  {author} {\bibinfo {author} {\bibfnamefont {V.}~\bibnamefont
  {Sanchez-Villicana}}, \bibinfo {author} {\bibfnamefont {S.~D.}\ \bibnamefont
  {Gensemer}}, \bibinfo {author} {\bibfnamefont {K.~Y.~N.}\ \bibnamefont
  {Tan}}, \bibinfo {author} {\bibfnamefont {A.}~\bibnamefont {Kumarakrishnan}},
  \bibinfo {author} {\bibfnamefont {T.~P.}\ \bibnamefont {Dinneen}}, \bibinfo
  {author} {\bibfnamefont {W.}~\bibnamefont {S\"uptitz}}, \ and\ \bibinfo
  {author} {\bibfnamefont {P.~L.}\ \bibnamefont {Gould}},\ }\href {\doibase
  10.1103/PhysRevLett.74.4619} {\bibfield  {journal} {\bibinfo  {journal}
  {Phys. Rev. Lett.}\ }\textbf {\bibinfo {volume} {74}},\ \bibinfo {pages}
  {4619} (\bibinfo {year} {1995})}\BibitemShut {NoStop}%
\bibitem [{\citenamefont {Suominen}\ \emph {et~al.}(1995)\citenamefont
  {Suominen}, \citenamefont {Holland}, \citenamefont {Burnett},\ and\
  \citenamefont {Julienne}}]{suominen95}%
  \BibitemOpen
  \bibfield  {author} {\bibinfo {author} {\bibfnamefont {K.-A.}\ \bibnamefont
  {Suominen}}, \bibinfo {author} {\bibfnamefont {M.~J.}\ \bibnamefont
  {Holland}}, \bibinfo {author} {\bibfnamefont {K.}~\bibnamefont {Burnett}}, \
  and\ \bibinfo {author} {\bibfnamefont {P.}~\bibnamefont {Julienne}},\ }\href
  {\doibase 10.1103/PhysRevA.51.1446} {\bibfield  {journal} {\bibinfo
  {journal} {Phys. Rev. A}\ }\textbf {\bibinfo {volume} {51}},\ \bibinfo
  {pages} {1446} (\bibinfo {year} {1995})}\BibitemShut {NoStop}%
\bibitem [{\citenamefont {Kraft}\ \emph {et~al.}(2005)\citenamefont {Kraft},
  \citenamefont {Mudrich}, \citenamefont {Staudt}, \citenamefont {Lange},
  \citenamefont {Dulieu}, \citenamefont {Wester},\ and\ \citenamefont
  {Weidem\"uller}}]{kraft05}%
  \BibitemOpen
  \bibfield  {author} {\bibinfo {author} {\bibfnamefont {S.~D.}\ \bibnamefont
  {Kraft}}, \bibinfo {author} {\bibfnamefont {M.}~\bibnamefont {Mudrich}},
  \bibinfo {author} {\bibfnamefont {M.~U.}\ \bibnamefont {Staudt}}, \bibinfo
  {author} {\bibfnamefont {J.}~\bibnamefont {Lange}}, \bibinfo {author}
  {\bibfnamefont {O.}~\bibnamefont {Dulieu}}, \bibinfo {author} {\bibfnamefont
  {R.}~\bibnamefont {Wester}}, \ and\ \bibinfo {author} {\bibfnamefont
  {M.}~\bibnamefont {Weidem\"uller}},\ }\href {\doibase
  10.1103/PhysRevA.71.013417} {\bibfield  {journal} {\bibinfo  {journal} {Phys.
  Rev. A}\ }\textbf {\bibinfo {volume} {71}},\ \bibinfo {pages} {013417}
  (\bibinfo {year} {2005})}\BibitemShut {NoStop}%
\bibitem [{\citenamefont {Boesten}\ \emph {et~al.}(1997)\citenamefont
  {Boesten}, \citenamefont {Tsai}, \citenamefont {Gardner}, \citenamefont
  {Heinzen},\ and\ \citenamefont {Verhaar}}]{boesten97}%
  \BibitemOpen
  \bibfield  {author} {\bibinfo {author} {\bibfnamefont {H.~M. J.~M.}\
  \bibnamefont {Boesten}}, \bibinfo {author} {\bibfnamefont {C.~C.}\
  \bibnamefont {Tsai}}, \bibinfo {author} {\bibfnamefont {J.~R.}\ \bibnamefont
  {Gardner}}, \bibinfo {author} {\bibfnamefont {D.~J.}\ \bibnamefont
  {Heinzen}}, \ and\ \bibinfo {author} {\bibfnamefont {B.~J.}\ \bibnamefont
  {Verhaar}},\ }\href {\doibase 10.1103/PhysRevA.55.636} {\bibfield  {journal}
  {\bibinfo  {journal} {Phys. Rev. A}\ }\textbf {\bibinfo {volume} {55}},\
  \bibinfo {pages} {636} (\bibinfo {year} {1997})}\BibitemShut {NoStop}%
\bibitem [{\citenamefont {Simoni}\ \emph {et~al.}(2002)\citenamefont {Simoni},
  \citenamefont {Julienne}, \citenamefont {Tiesinga},\ and\ \citenamefont
  {Williams}}]{simoni02}%
  \BibitemOpen
  \bibfield  {author} {\bibinfo {author} {\bibfnamefont {A.}~\bibnamefont
  {Simoni}}, \bibinfo {author} {\bibfnamefont {P.~S.}\ \bibnamefont
  {Julienne}}, \bibinfo {author} {\bibfnamefont {E.}~\bibnamefont {Tiesinga}},
  \ and\ \bibinfo {author} {\bibfnamefont {C.~J.}\ \bibnamefont {Williams}},\
  }\href {\doibase 10.1103/PhysRevA.66.063406} {\bibfield  {journal} {\bibinfo
  {journal} {Phys. Rev. A}\ }\textbf {\bibinfo {volume} {66}},\ \bibinfo
  {pages} {063406} (\bibinfo {year} {2002})}\BibitemShut {NoStop}%
\bibitem [{\citenamefont {Volz}\ \emph {et~al.}(2005)\citenamefont {Volz},
  \citenamefont {D\"urr}, \citenamefont {Syassen}, \citenamefont {Rempe},
  \citenamefont {van Kempen},\ and\ \citenamefont {Kokkelmans}}]{voltz05}%
  \BibitemOpen
  \bibfield  {author} {\bibinfo {author} {\bibfnamefont {T.}~\bibnamefont
  {Volz}}, \bibinfo {author} {\bibfnamefont {S.}~\bibnamefont {D\"urr}},
  \bibinfo {author} {\bibfnamefont {N.}~\bibnamefont {Syassen}}, \bibinfo
  {author} {\bibfnamefont {G.}~\bibnamefont {Rempe}}, \bibinfo {author}
  {\bibfnamefont {E.}~\bibnamefont {van Kempen}}, \ and\ \bibinfo {author}
  {\bibfnamefont {S.}~\bibnamefont {Kokkelmans}},\ }\href {\doibase
  10.1103/PhysRevA.72.010704} {\bibfield  {journal} {\bibinfo  {journal} {Phys.
  Rev. A}\ }\textbf {\bibinfo {volume} {72}},\ \bibinfo {pages} {010704}
  (\bibinfo {year} {2005})}\BibitemShut {NoStop}%
\bibitem [{\citenamefont {Buggle}\ \emph {et~al.}(2004)\citenamefont {Buggle},
  \citenamefont {L\'eonard}, \citenamefont {von Klitzing},\ and\ \citenamefont
  {Walraven}}]{buggle04}%
  \BibitemOpen
  \bibfield  {author} {\bibinfo {author} {\bibfnamefont {C.}~\bibnamefont
  {Buggle}}, \bibinfo {author} {\bibfnamefont {J.}~\bibnamefont {L\'eonard}},
  \bibinfo {author} {\bibfnamefont {W.}~\bibnamefont {von Klitzing}}, \ and\
  \bibinfo {author} {\bibfnamefont {J.~T.~M.}\ \bibnamefont {Walraven}},\
  }\href {\doibase 10.1103/PhysRevLett.93.173202} {\bibfield  {journal}
  {\bibinfo  {journal} {Phys. Rev. Lett.}\ }\textbf {\bibinfo {volume} {93}},\
  \bibinfo {pages} {173202} (\bibinfo {year} {2004})}\BibitemShut {NoStop}%
\bibitem [{\citenamefont {Jones}\ \emph {et~al.}(1999)\citenamefont {Jones},
  \citenamefont {Lett}, \citenamefont {Tiesinga},\ and\ \citenamefont
  {Julienne}}]{jones99}%
  \BibitemOpen
  \bibfield  {author} {\bibinfo {author} {\bibfnamefont {K.~M.}\ \bibnamefont
  {Jones}}, \bibinfo {author} {\bibfnamefont {P.~D.}\ \bibnamefont {Lett}},
  \bibinfo {author} {\bibfnamefont {E.}~\bibnamefont {Tiesinga}}, \ and\
  \bibinfo {author} {\bibfnamefont {P.~S.}\ \bibnamefont {Julienne}},\ }\href
  {\doibase 10.1103/PhysRevA.61.012501} {\bibfield  {journal} {\bibinfo
  {journal} {Phys. Rev. A}\ }\textbf {\bibinfo {volume} {61}},\ \bibinfo
  {pages} {012501} (\bibinfo {year} {1999})}\BibitemShut {NoStop}%
\bibitem [{\citenamefont {Tiesenga}\ \emph {et~al.}(1996)\citenamefont
  {Tiesenga}, \citenamefont {Williams}, \citenamefont {Julienne}, \citenamefont
  {Jones}, \citenamefont {Lett},\ and\ \citenamefont {Phillips}}]{tiesenga96}%
  \BibitemOpen
  \bibfield  {author} {\bibinfo {author} {\bibfnamefont {E.}~\bibnamefont
  {Tiesenga}}, \bibinfo {author} {\bibfnamefont {C.~J.}\ \bibnamefont
  {Williams}}, \bibinfo {author} {\bibfnamefont {P.~S.}\ \bibnamefont
  {Julienne}}, \bibinfo {author} {\bibfnamefont {K.~M.}\ \bibnamefont {Jones}},
  \bibinfo {author} {\bibfnamefont {P.~D.}\ \bibnamefont {Lett}}, \ and\
  \bibinfo {author} {\bibfnamefont {W.~D.}\ \bibnamefont {Phillips}},\ }\href
  {\doibase 10.6028/jres.101.051} {\bibfield  {journal} {\bibinfo  {journal}
  {J. Res. Natl. Inst. Stand. Technol.}\ }\textbf {\bibinfo {volume} {101}},\
  \bibinfo {pages} {505} (\bibinfo {year} {1996})}\BibitemShut {NoStop}%
\bibitem [{\citenamefont {Londo\~no}\ \emph {et~al.}(2010)\citenamefont
  {Londo\~no}, \citenamefont {Mahecha}, \citenamefont {Luc-Koenig},\ and\
  \citenamefont {Crubellier}}]{londono10}%
  \BibitemOpen
  \bibfield  {author} {\bibinfo {author} {\bibfnamefont {B.~E.}\ \bibnamefont
  {Londo\~no}}, \bibinfo {author} {\bibfnamefont {J.~E.}\ \bibnamefont
  {Mahecha}}, \bibinfo {author} {\bibfnamefont {E.}~\bibnamefont {Luc-Koenig}},
  \ and\ \bibinfo {author} {\bibfnamefont {A.}~\bibnamefont {Crubellier}},\
  }\href {\doibase 10.1103/PhysRevA.82.012510} {\bibfield  {journal} {\bibinfo
  {journal} {Phys. Rev. A}\ }\textbf {\bibinfo {volume} {82}},\ \bibinfo
  {pages} {012510} (\bibinfo {year} {2010})}\BibitemShut {NoStop}%
\bibitem [{\citenamefont {Thomas}\ \emph {et~al.}(2004)\citenamefont {Thomas},
  \citenamefont {Kj\ae{}rgaard}, \citenamefont {Julienne},\ and\ \citenamefont
  {Wilson}}]{thomas04}%
  \BibitemOpen
  \bibfield  {author} {\bibinfo {author} {\bibfnamefont {N.~R.}\ \bibnamefont
  {Thomas}}, \bibinfo {author} {\bibfnamefont {N.}~\bibnamefont
  {Kj\ae{}rgaard}}, \bibinfo {author} {\bibfnamefont {P.~S.}\ \bibnamefont
  {Julienne}}, \ and\ \bibinfo {author} {\bibfnamefont {A.~C.}\ \bibnamefont
  {Wilson}},\ }\href {\doibase 10.1103/PhysRevLett.93.173201} {\bibfield
  {journal} {\bibinfo  {journal} {Phys. Rev. Lett.}\ }\textbf {\bibinfo
  {volume} {93}},\ \bibinfo {pages} {173201} (\bibinfo {year}
  {2004})}\BibitemShut {NoStop}%
\bibitem [{\citenamefont {DeMarco}\ \emph {et~al.}(1999)\citenamefont
  {DeMarco}, \citenamefont {Bohn}, \citenamefont {Burke}, \citenamefont
  {Holland},\ and\ \citenamefont {Jin}}]{demarco99}%
  \BibitemOpen
  \bibfield  {author} {\bibinfo {author} {\bibfnamefont {B.}~\bibnamefont
  {DeMarco}}, \bibinfo {author} {\bibfnamefont {J.~L.}\ \bibnamefont {Bohn}},
  \bibinfo {author} {\bibfnamefont {J.~P.}\ \bibnamefont {Burke}}, \bibinfo
  {author} {\bibfnamefont {M.}~\bibnamefont {Holland}}, \ and\ \bibinfo
  {author} {\bibfnamefont {D.~S.}\ \bibnamefont {Jin}},\ }\href {\doibase
  10.1103/PhysRevLett.82.4208} {\bibfield  {journal} {\bibinfo  {journal}
  {Phys. Rev. Lett.}\ }\textbf {\bibinfo {volume} {82}},\ \bibinfo {pages}
  {4208} (\bibinfo {year} {1999})}\BibitemShut {NoStop}%
\bibitem [{\citenamefont {Burke}\ \emph {et~al.}(1999)\citenamefont {Burke},
  \citenamefont {Greene}, \citenamefont {Bohn}, \citenamefont {Wang},
  \citenamefont {Gould},\ and\ \citenamefont {Stwalley}}]{burke99}%
  \BibitemOpen
  \bibfield  {author} {\bibinfo {author} {\bibfnamefont {J.~P.}\ \bibnamefont
  {Burke}}, \bibinfo {author} {\bibfnamefont {C.~H.}\ \bibnamefont {Greene}},
  \bibinfo {author} {\bibfnamefont {J.~L.}\ \bibnamefont {Bohn}}, \bibinfo
  {author} {\bibfnamefont {H.}~\bibnamefont {Wang}}, \bibinfo {author}
  {\bibfnamefont {P.~L.}\ \bibnamefont {Gould}}, \ and\ \bibinfo {author}
  {\bibfnamefont {W.~C.}\ \bibnamefont {Stwalley}},\ }\href {\doibase
  10.1103/PhysRevA.60.4417} {\bibfield  {journal} {\bibinfo  {journal} {Phys.
  Rev. A}\ }\textbf {\bibinfo {volume} {60}},\ \bibinfo {pages} {4417}
  (\bibinfo {year} {1999})}\BibitemShut {NoStop}%
\bibitem [{\citenamefont {C\^ot\'e}\ \emph {et~al.}(1999)\citenamefont
  {C\^ot\'e}, \citenamefont {Dalgarno}, \citenamefont {Lyyra},\ and\
  \citenamefont {Li}}]{cote99}%
  \BibitemOpen
  \bibfield  {author} {\bibinfo {author} {\bibfnamefont {R.}~\bibnamefont
  {C\^ot\'e}}, \bibinfo {author} {\bibfnamefont {A.}~\bibnamefont {Dalgarno}},
  \bibinfo {author} {\bibfnamefont {A.~M.}\ \bibnamefont {Lyyra}}, \ and\
  \bibinfo {author} {\bibfnamefont {L.}~\bibnamefont {Li}},\ }\href {\doibase
  10.1103/PhysRevA.60.2063} {\bibfield  {journal} {\bibinfo  {journal} {Phys.
  Rev. A}\ }\textbf {\bibinfo {volume} {60}},\ \bibinfo {pages} {2063}
  (\bibinfo {year} {1999})}\BibitemShut {NoStop}%
\bibitem [{\citenamefont {Fioretti}\ \emph {et~al.}(1999)\citenamefont
  {Fioretti}, \citenamefont {Comparat}, \citenamefont {Drag}, \citenamefont
  {Gallagher},\ and\ \citenamefont {Pillet}}]{fioretti99}%
  \BibitemOpen
  \bibfield  {author} {\bibinfo {author} {\bibfnamefont {A.}~\bibnamefont
  {Fioretti}}, \bibinfo {author} {\bibfnamefont {D.}~\bibnamefont {Comparat}},
  \bibinfo {author} {\bibfnamefont {C.}~\bibnamefont {Drag}}, \bibinfo {author}
  {\bibfnamefont {T.~F.}\ \bibnamefont {Gallagher}}, \ and\ \bibinfo {author}
  {\bibfnamefont {P.}~\bibnamefont {Pillet}},\ }\href {\doibase
  10.1103/PhysRevLett.82.1839} {\bibfield  {journal} {\bibinfo  {journal}
  {Phys. Rev. Lett.}\ }\textbf {\bibinfo {volume} {82}},\ \bibinfo {pages}
  {1839} (\bibinfo {year} {1999})}\BibitemShut {NoStop}%
\bibitem [{\citenamefont {Shaffer}\ \emph {et~al.}(1999)\citenamefont
  {Shaffer}, \citenamefont {Chalupczak},\ and\ \citenamefont
  {Bigelow}}]{shaffer99}%
  \BibitemOpen
  \bibfield  {author} {\bibinfo {author} {\bibfnamefont {J.~P.}\ \bibnamefont
  {Shaffer}}, \bibinfo {author} {\bibfnamefont {W.}~\bibnamefont {Chalupczak}},
  \ and\ \bibinfo {author} {\bibfnamefont {N.~P.}\ \bibnamefont {Bigelow}},\
  }\href {\doibase 10.1103/PhysRevLett.83.3621} {\bibfield  {journal} {\bibinfo
   {journal} {Phys. Rev. Lett.}\ }\textbf {\bibinfo {volume} {83}},\ \bibinfo
  {pages} {3621} (\bibinfo {year} {1999})}\BibitemShut {NoStop}%
\bibitem [{\citenamefont {Gomez}\ \emph {et~al.}(2007)\citenamefont {Gomez},
  \citenamefont {Black}, \citenamefont {Turner}, \citenamefont {Tiesinga},\
  and\ \citenamefont {Lett}}]{gomez07}%
  \BibitemOpen
  \bibfield  {author} {\bibinfo {author} {\bibfnamefont {E.}~\bibnamefont
  {Gomez}}, \bibinfo {author} {\bibfnamefont {A.~T.}\ \bibnamefont {Black}},
  \bibinfo {author} {\bibfnamefont {L.~D.}\ \bibnamefont {Turner}}, \bibinfo
  {author} {\bibfnamefont {E.}~\bibnamefont {Tiesinga}}, \ and\ \bibinfo
  {author} {\bibfnamefont {P.~D.}\ \bibnamefont {Lett}},\ }\href {\doibase
  10.1103/PhysRevA.75.013420} {\bibfield  {journal} {\bibinfo  {journal} {Phys.
  Rev. A}\ }\textbf {\bibinfo {volume} {75}},\ \bibinfo {pages} {013420}
  (\bibinfo {year} {2007})}\BibitemShut {NoStop}%
\end{thebibliography}%

\end{document}